\documentclass[a4paper,11pt]{article}
\pdfoutput=1 
\usepackage{jheppub} 
\usepackage[T1]{fontenc} 
\usepackage{amsmath,amssymb,epsfig,txfonts,relsize,slashed}
\usepackage[utf8]{inputenc}

\allowdisplaybreaks

\definecolor{nicered}{rgb}{0.7,0.1,0.1}
\definecolor{nicegreen}{rgb}{0.1,0.5,0.1}
\definecolor{niceblue}{rgb}{0.0,0.1,0.7}
\hypersetup{colorlinks,citecolor=nicegreen,linkcolor=nicered,urlcolor=niceblue}

\def \beq{\begin{equation}}
\def \eeq{\end{equation}}
\def \bea{\begin{eqnarray}}
\def \eea{\end{eqnarray}}

\title{Bounds on CP-violating Higgs-gluon interactions: \\ the case of vanishing light-quark Yukawa couplings}

\author[1]{Ulrich Haisch}

\author[1]{and Amando Hala}

\affiliation[1]{Max Planck Institute for Physics, F{\"o}hringer Ring 6,  80805 M{\"u}nchen, Germany}   

\emailAdd{haisch@mpp.mpg.de}
\emailAdd{ahala@mpp.mpg.de}

\abstract{
\phantom{iii} We investigate CP-violating interactions involving the Higgs boson and gluons within an effective field theory approach, focusing on the specific class of new-physics scenarios where the Yukawa couplings of light quarks are zero or strongly suppressed compared to the standard-model expectations. We compute the contributions of the most relevant higher-dimensional operators of Weinberg type to the electric dipole moment of the neutron~(nEDM), which are induced by Feynman diagrams that involve an effective CP-violating Higgs-gluon coupling and top-quark loops. The  resulting nEDM sensitivities and prospects are discussed and compared to the existing and expected LHC bounds. We find that future nEDM searches can set non-trivial constraints on  CP-violating Higgs-gluon interactions even if the Higgs only couples to the third generation of quarks.}

\preprint{}

\begin{document} 

\maketitle

\section{Motivation}
\label{sec:motivation}

It is known since more than a decade~\cite{Plehn:2001nj,Klamke:2007cu} that measurements of the kinematic properties of the Higgs boson and the associated jet spectra provide experimental probes of the CP structure of the couplings between the Higgs and gauge bosons. In the context of the standard model effective field theory~(SMEFT)~\cite{Buchmuller:1985jz,Grzadkowski:2010es} one  CP-violating dimension-six operator, that has  been constrained using LHC data~\cite{Aad:2015tna,Ferreira:2016jea,Aaboud:2018xdt,Bernlochner:2018opw}, is 
\beq \label{eq:LCPVexample}
{\cal L}_{\phi \tilde G} = -g_s^2 \, \phi^\dagger \phi \, \tilde G_{\mu \nu}^a  G^{a \, \mu \nu} \, C_{\phi \tilde G} \,.
\eeq
Here $g_s$ denotes the strong coupling constant, $\phi$ is the standard model (SM) Higgs doublet, $G^a_{\mu \nu}$ is the QCD field strength tensor and $\tilde G^{a \, {\mu \nu}} = 1/2 \hspace{0.5mm} \epsilon^{\mu \nu \rho \lambda} \, G^a_{\rho \lambda}$ with $\epsilon^{0123} = +1$ is its dual. Notice finally that the Wilson coefficient $C_{\phi \tilde G}$ introduced in~(\ref{eq:LCPVexample}) carries mass dimension $-2$, meaning that it can be measured in~$1/{\rm GeV}^{2}$.

Since  searches for electric dipole moments (EDMs) are also known to place stringent constraints on any new-physics scenario with additional sources of CP violation (see~\cite{Pospelov:2005pr,Li:2010ax,Engel:2013lsa,Jung:2013hka,Brod:2013cka,Inoue:2014nva,Gorbahn:2014sha,Altmannshofer:2015qra,Chien:2015xha,Cirigliano:2016njn,Cirigliano:2016nyn,Yamanaka:2017mef,Yanase:2018qqq,Dekens:2018bci,Brod:2018pli,Brod:2018lbf} for reviews and recent discussions) the low-energy constraints on effective operators of the form~(\ref{eq:LCPVexample}) have also been considered~\cite{McKeen:2012av,Chang:2013cia,Gripaios:2013lea,Dwivedi:2015nta,Cesarotti:2018huy,Panico:2018hal,Cirigliano:2019vfc}. In fact, the recent article~\cite{Cirigliano:2019vfc} performed a comprehensive study of the relative strengths and complementarity of collider and low-energy measurements in probing CP-violation in Higgs-gauge boson interactions. Employing a SMEFT description and working in the context of so-called universal theories~\cite{Barbieri:2004qk,Gripaios:2009pe,Espinosa:2011eu},~i.e.~theories in which mainly the couplings between the  SM Higgs and gauge bosons are modified by new dynamics, it was found in the latter work that in a single-operator analysis the existing EDM limits leave very little room for observing CP violation in the Higgs sector at the LHC.  Including all relevant dimension-six CP-violating operators, it was furthermore established  that the EDM searches  enforce strong correlations among Higgs-gauge boson couplings, which barring intricate cancellations lead again to stringent  bounds on the individual Wilson coefficients. Similar conclusions where drawn in~\cite{Panico:2018hal} where only the limits arising from the electron EDM (eEDM) have been studied. 

In this article we would like to point out a simple way of how-to relax the constraints obtained in~\cite{Panico:2018hal,Cirigliano:2019vfc}. In contrast to the latter articles, we will not assume that the new-physics modifications are confined to the Higgs-gauge boson sector, but also allow for effects in the Yukawa sector. Specifically, we will consider the following dimension-six SMEFT terms 
\bea \label{eq:LCPQCD}
{\cal L}_{\phi q}   =  -Y_d \, \phi^\dagger \phi  \,  \bar Q_L  \hspace{0.25mm}  \phi \hspace{0.25mm}  d_R \, C_{\phi d}  -Y_u \, \phi^\dagger \phi   \, \bar Q_L \hspace{0.25mm}  \tilde \phi \hspace{0.25mm} u_R  \, C_{\phi u} + {\rm h.c.} \,,
\eea
where we have employed the shorthand notation $\tilde \phi^i = \epsilon_{ij} \hspace{0.5mm} \big ( \phi^j \big )^\ast$ with $\epsilon_{ij}$ totally anti-symmetric and
$\epsilon_{12} = +1$. The Yukawa couplings $Y_d$ and $Y_u$ are matrices in flavour space and a sum over flavours indices is implicit in~(\ref{eq:LCPQCD}). Finally, $Q_L$ denote left-handed quark doublets, while $d_R$ and $u_R$ are right-handed fermion singlets of down-quark and up-quark type, respectively.

After electroweak symmetry breaking~(EWSB) the dimension-six operators in~(\ref{eq:LCPQCD}) modify the couplings of the Higgs boson to quarks. Assuming that new-physics is minimally flavour violating~\cite{DAmbrosio:2002vsn} and that the Wilson coefficients $C_{\phi u}$ and $C_{\phi d}$ are real,\footnote{CP-violating diagonal~\cite{Brod:2013cka,Brod:2018pli} and flavour-changing Higgs-fermion~\cite{Gorbahn:2014sha} couplings involving the third generation would be subject to stringent EDM constraints.} each SM quark Yukawa coupling gets rescaled by an independent factor 
\beq \label{eq:kappaq}
\kappa_q \simeq 1 + v^2 \, C_{\phi q} \,, 
\eeq
where $q = t,b,c,s,d,u$ and $v \simeq 246 \, {\rm GeV}$ denotes the electroweak~(EW) vacuum expectation value. The coupling modifiers $\kappa_q$ or equivalently the Wilson coefficients $C_{\phi q}$ can be constrained by LHC Higgs physics. In the case of the top and bottom quark, our knowledge of Yukawa interactions has undergone a revolution in the last year, since the ATLAS and CMS collaborations have independently observed $pp \to t \bar t h$ production~\cite{Sirunyan:2018hoz,Aaboud:2018urx} and the $h \to b \bar b$ decay~\cite{Aaboud:2018zhk,Sirunyan:2018kst}. Combining direct and indirect information on the Higgs properties into a global fit ATLAS~\cite{ATLAS-CONF-2019-005} finds the following 68\%~confidence level~(CL) limits\footnote{In the considered benchmark model no new-physics contributions to Higgs-boson decays are assumed to exist and Higgs-boson vertices involving loops are resolved in terms of their SM content.}
\beq \label{eq:kappabkappat}  
\kappa_t = 1.02^{+0.11}_{-0.10} \,, \qquad \kappa_b = 1.06^{+0.19}_{-0.18} \,.
\eeq
The quoted results  are in full agreement with the bounds obtained by CMS~\cite{Sirunyan:2018koj}, and yield clear and model-independent evidence for the existence of non-zero top-quark  and bottom-quark Yukawa couplings in nature. Despite significant experimental  and theoretical effort~\cite{Bodwin:2013gca,Delaunay:2013pja,Kagan:2014ila,Aad:2015sda,Perez:2015aoa,Koenig:2015pha,Perez:2015lra,Brivio:2015fxa,Bishara:2016jga,Soreq:2016rae,Yu:2016rvv,Han:2017yhy,Cohen:2017rsk,Aaboud:2018fhh,CMS-PAS-FTR-18-011,Sirunyan:2018sgc,Han:2018juw,ATLAS-CONF-2019-029,CMS-PAS-HIG-18-031,Sirunyan:2019lhe,Alasfar:2019pmn} only very weak (no relevant) bounds exist at present in the case of the second-generation (first-generation) quarks.  Whether the Higgs mechanism is responsible for the generation of the masses of the charm, strange, down and up quark is thus an open question, and new-physics scenarios~(see~e.g.~\cite{Ghosh:2015gpa})  that predict a significant reduction of the couplings of the observed Higgs boson to the first two generation of quarks,~i.e.~$\kappa_{c,s,d,u} \simeq 0$, are from the phenomenological point of view a viable option. 

\begin{figure}[!t]
\begin{center}
\includegraphics[width=\textwidth]{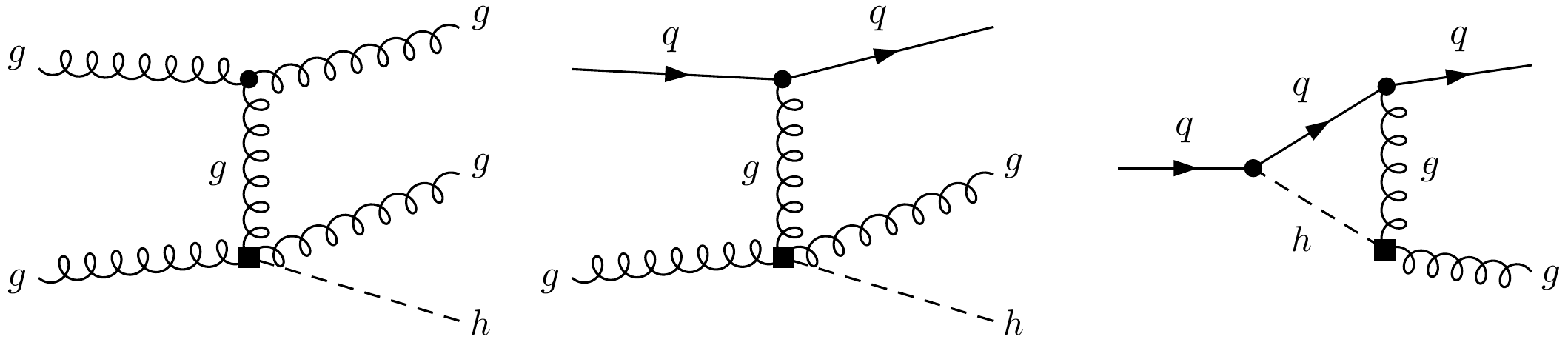} 
\vspace{-2mm}
\caption{\label{fig:diagrams1}   Left: Tree-level graphs that give rise to $pp \to h + 2j$ production at the LHC. Right:  Example one-loop diagram that contributes for instance to the nEDM. The black squares indicate insertions of the CP-violating dimension-six operator~(\ref{eq:LCPVexample}).} 
\end{center}
\end{figure}

A question that one therefore may want to ask is how sensitively the LHC limits~\cite{Aad:2015tna,Ferreira:2016jea,Aaboud:2018xdt,Bernlochner:2018opw} and the~EDM constraints~\cite{Panico:2018hal,Cirigliano:2019vfc} depend on the assumption that the observed Higgs boson has couplings to the light fermions. In the case of the operator~(\ref{eq:LCPVexample}) this question can be answered immediately by looking at the Feynman diagrams depicted in Figure~\ref{fig:diagrams1}. The two tree-level diagrams on the left-hand side give a correction to Higgs plus two jet production $pp\to h +2j$, while the one-loop graph shown on the  right induces chromoelectric dipole moments~(CEDMs) for light quarks, which in turn generate non-zero contributions to the neutron EDM~(nEDM) and all other hadronic~EDMs. Since the diagrams that lead to the LHC signal do not involve a vertex where the Higgs couples to a light quark, the constraints on $C_{\phi \tilde G}$ that can be obtained from kinematic properties of $pp\to h +2j$ are obviously independent of the size of the $\kappa_q$ parameters. The amplitude of the chromoelectric dipole transition $q \to q g$ instead depends  linearly on $\kappa_q$, and hence tends to zero in the limit of vanishing light-quark Yukawa couplings. In consequence,  if the down-quark and up-quark Yukawa couplings are identical to zero, no bound on the Wilson coefficient~$C_{\phi \tilde G}$ can be obtained from hadronic EDM searches at the one-loop level. The same statement can be shown to hold for the Wilson coefficients of the dimension-six operators which encode the CP-violating couplings between the Higgs and EW gauge bosons and contribute to the eEDM. Since in~\cite{Panico:2018hal,Cirigliano:2019vfc} it is assumed  that the Yukawa couplings of light fermions are exactly SM-like, it follows that the limits derived in these papers do not directly apply in the case that the light-fermion Yukawa couplings vanish exactly or are strongly suppressed. 

Motivated by the above observation, we compute in Section~\ref{sec:calculation} of this article the leading contributions to the nEDM that involve an insertion of~(\ref{eq:LCPVexample}) and that survive in the limit of vanishing light-quark Yukawa couplings. An extension of our calculation to the case of the full set of dimension-six operators that encode the CP-violating interactions between the Higgs and EW gauge bosons will be presented elsewhere~\cite{inprep}. Based on our results, we then derive in Section~\ref{sec:discussion} the bounds on the Wilson coefficient of the CP-violating dimension-six operator~(\ref{eq:LCPVexample}) from current and future nEDM searches, comparing our findings to the existing LHC limits and their projections.  In Appendix~\ref{sec:SMyukawas} we extend the formalism presented in Section~\ref{sec:dim6} to the case where the light-quark Yukawa couplings are SM-like.

\section{Calculation}
\label{sec:calculation}

Before describing the basic steps of our calculation, we mention that after EWSB the operator~(\ref{eq:LCPVexample}) shifts the QCD $\theta$ term ${\cal L}_\theta = g_s^2/(32 \hspace{0.125mm} \pi^2) \,\theta \, \tilde G_{\mu \nu}^a  G^{a \, \mu \nu} $ by a finite amount,~i.e.~$\theta \to \theta - 16 \hspace{0.25mm} \pi^2 \hspace{0.25mm} v^2 \, C_{\phi \tilde G}$. Since  based on the current experimental nEDM results~\cite{Baker:2006ts,Afach:2015sja} one has $\left |\theta \right | \lesssim 10^{-10}$, we will assume that the total $\theta$ term vanishes dynamically due to a Peccei-Quinn mechanism~\cite{Peccei:1977hh}.  Under this assumption there is no direct bound on~(\ref{eq:LCPVexample}), and the Wilson coefficient $C_{\phi \tilde G}$ can be treated as a free parameter in the SMEFT as done  in the analyses~\cite{Aad:2015tna,Ferreira:2016jea,Aaboud:2018xdt,Bernlochner:2018opw,Cirigliano:2019vfc}.  

As explained at the end of Section~\ref{sec:motivation}, the goal of this work is it to calculate the numerically most relevant contributions to the nEDM that are proportional to the Wilson coefficient $C_{\phi \tilde G}$ and that involve the Yukawa couplings of the third-generation quarks. It turns out that at the matching scale the relevant loop graphs give rise to the following two CP-violating higher-dimensional operators of Weinberg~type~\cite{Weinberg:1989dx,Morozov:1985ef,Chang:1991ry,Booth:1992iz}
\beq \label{eq:LCPVQCD}
{\cal L}_{W1} =  -\frac{g_s}{3} \hspace{0.25mm} f^{abc} \hspace{0.5mm}  \tilde G_{\mu \nu}^a G^{b \, \nu \rho}  G^{c \, \mu}_{\rho}  \, C_{3 \tilde G} - \frac{g_s^2}{12} \,  \tilde G_{\mu \nu}^a G^{a \, \mu \nu}  G_{\rho \lambda}^b G^{b \, \rho \lambda}  \, C_{4 \tilde G, 1} \,,
\eeq
where $f^{abc}$ are the fully anti-symmetric structure constants of $S\!U(3)$ colour. Since the bottom-quark contributions can be shown to be suppressed relative to the top-quark effects by a factor of~$m_b^2/m_h^2 \hspace{0.25mm} \ln^2 \left (m_b^2/m_h^2 \right ) \simeq 5\%$, we neglect corrections that are proportional to the bottom-quark Yukawa coupling in what follows. The top-quark effects are all linearly dependent on the coupling modifier~$\kappa_t$. In view of the observed SM-like nature~(\ref{eq:kappabkappat}) of the top-quark Yukawa coupling, we will simply employ~$\kappa_t =1$ in our calculations. Allowing for ${\cal O} (10\%)$ variations of $\kappa_t$ would, however, not qualitatively change the results of our numerical analysis performed in Section~\ref{sec:discussion}. 

\subsection{Dimension-six contribution}
\label{sec:dim6}

\begin{figure}[!t]
\begin{center}
\includegraphics[width=\textwidth]{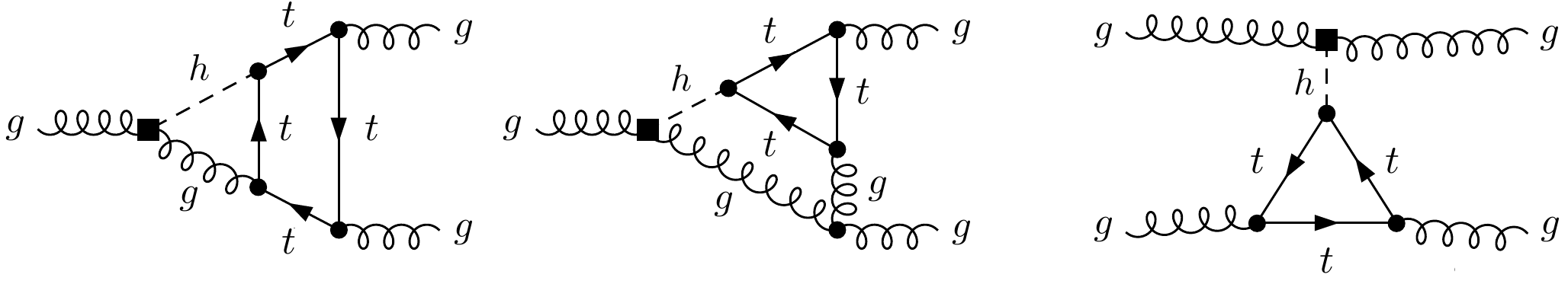} 
\vspace{-6mm}
\caption{\label{fig:diagrams2} Left:  Example diagrams of two-loop corrections to $C_{3 \tilde G}$ arising from the insertion of $C_{\phi \tilde G}$.  Right:~A one-loop correction to $C_{4 \tilde G, 1}$ arising from the insertion of $C_{\phi \tilde G}$. The operator insertions are indicated by black squares.} 
\end{center}
\end{figure}

The leading-order (LO) matching correction to the Wilson coefficient $C_{3 \tilde G}$ proportional to  $C_{\phi \tilde G}$ arises from two-loop Feynman diagrams  like the ones displayed on the left-hand side in Figure~\ref{fig:diagrams2}. Employing a hard mass procedure (see~\cite{Smirnov:2002pj} for a review) to obtain systematic expansions of the relevant two-loop diagrams  in powers of the external momenta and the ratio $x = m_t^2/m_h^2$ with $m_t \simeq 163 \, {\rm GeV}$ and $m_h \simeq 125 \, {\rm GeV}$ the top-quark and Higgs-boson mass,  we find the following analytic result
\beq \label{eq:twomatch} 
C_{3 \tilde G} (m_h) = \frac{ \alpha_s^2 (m_h)}{8 \hspace{0.125mm} \pi^2} \, \left [ 
\frac{65}{6} + 2 \ln x 
+ \frac{1}{x} \left ( \frac{383}{900} + \frac{2}{15} \ln x \right ) 
\right ]  C_{\phi \tilde G} (m_h) \,,
\eeq 
where $\alpha_s = g_s^2/(4 \hspace{0.125mm} \pi)$. The expression given above corresponds to the  $\overline{\rm MS}$ scheme with the renormalisation scale set to $\mu = m_h$.\footnote{The sum of the bare two-loop Feynman diagrams that contributes to~(\ref{eq:twomatch}) is not ultraviolet~(UV)~finite. The remaining~UV pole is cancelled by taking into account  the one-loop mixing of the operator $ \phi^\dagger \phi \, G_{\mu \nu}^a \tilde G^{a \, \mu \nu} $ into $\bar Q_L \hspace{0.25mm} \sigma^{\mu \nu} t^a \hspace{0.25mm} u_R \, \tilde \phi \, G_{\mu \nu}^a + {\rm h.c.}$ with  $t^a$ denoting the generator of $S\!U(3)$ colour. We have calculated the relevant one-loop mixing finding agreement with the result  given in~\cite{Alonso:2013hga}.} The actual calculation was performed in a $R_\xi$ background field gauge for the gluon~\cite{Abbott:1980hw,Abbott:1981ke} keeping an arbitrary gauge parameter. The Levi-Civita tensor $\epsilon_{\mu \nu \rho \lambda}$ was treated as an external four-dimensional object. Our computations made use of the in-house codes that were developed  in the context of~\cite{Bizon:2018syu,Gorbahn:2019lwq}, except for the  tensor reduction of two-point and three-point one-loop integrals which relied on~{\tt Package-X}~\cite{Patel:2015tea}. We add that we have calculated higher-order terms in the $1/x$ expansion of $C_{3 \tilde G} (m_h)$ and found that these corrections shift  the numerical value of the  matching correction~(\ref{eq:twomatch}) by less than a permille.\footnote{The analytic expressions for the ${\cal O} (1/x^2)$ and ${\cal O} (1/x^3)$ terms can be found in the {\LaTeX} source code of this article.}  Such an accuracy is  more than sufficient for our purpose. 

The renormalisation group (RG) flow  from the EW  to the hadronic scale $\mu_H = 1 \, {\rm GeV}$ does not only change the  value of the Wilson coefficient $C_{3 \tilde G}$, but also induces non-zero contributions for the EDMs $d_q$ and the CEDMs $\tilde d_q$ of the down  and up quarks 
\beq \label{eq:calLd}
{\cal L}_d = -\frac{i}{2} \hspace{0.5mm} \bar q \hspace{0.25mm} \sigma_{\mu \nu} \gamma_5  \hspace{0.25mm}q  \hspace{0.25mm} F^{\mu \nu} \hspace{0.5mm} d_q  -\frac{i}{2} \hspace{0.5mm}  g_s \hspace{0.125mm} \bar q \hspace{0.25mm}  \sigma_{\mu \nu} \hspace{0.25mm}  t^a   \gamma_5  \hspace{0.25mm}q  \hspace{0.25mm} G^{a \, \mu \nu} \, \tilde d_q \,, 
\eeq
with $q = d, u$, $\sigma^{\mu \nu} = i/2 \left ( \gamma^\mu  \gamma^\nu - \gamma^\mu  \gamma^\nu \right )$, and $F_{\mu \nu}$ denotes the QED field strength tensor. In the basis $\vec C_6 = \big (d_q, \tilde d_q, C_{3 \tilde G} \big)^T$, the one-loop anomalous dimension~(AD) matrix takes the following form~\cite{Braaten:1990gq,Braaten:1990zt,Degrassi:2005zd}
\beq \label{eq:ADM3G} 
\hat \gamma_6 = \begin{pmatrix} \; \displaystyle \frac{32}{3}  \; & \; 0 \; &  \; 0 \; \\[3mm] \; \displaystyle \frac{32}{3}  \; & \; \displaystyle \frac{28}{3} \; &  \; 0\;  \\[3mm] \; 0\; & \;-6 \;& \;3 + 2 N_F + 2 \hspace{0.25mm} \beta_0 \; \end{pmatrix} \,, 
\eeq
where $\beta_0 = 11 - 2/3 \hspace{0.25mm} N_F$ is the LO QCD beta function and $N_F$ denotes the number of active quark flavours.  Resumming  leading-logarithmic~(LL) corrections in the five-flavour, four-flavour and three-flavour  theory, we obtain 
\beq \label{eq:3GRG} 
\begin{split}
d_q (\mu_H) & \simeq -5.6 \cdot 10^{-2} \,  e \hspace{0.25mm} Q_q  \hspace{0.25mm} m_q (\mu_H) \, C_{3 \tilde G} (m_h) \,, \\[2mm] 
\tilde d_q (\mu_H) &  \simeq 1.2 \cdot 10^{-1}  \, m_q (\mu_H) \, C_{3 \tilde G} (m_h)  \,,  \\[3mm] 
C_{3 \tilde G} (\mu_H) & \simeq 1.3 \cdot 10^{-1} \hspace{0.5mm}  C_{3 \tilde G} (m_h) \,.
\end{split}
\eeq
Here $e$ denotes the electron charge magnitude, $Q_q$ is the fractional electric charge of the relevant quark and $m_q (\mu_H)$ is its $\overline{\rm MS}$ mass at the hadronic scale.  The numerical  factors  in~(\ref{eq:3GRG}) correspond to  the  values $\alpha_s (m_h) \simeq 0.11$, $\alpha_s (m_b) \simeq 0.21$, $\alpha_s (m_c) \simeq 0.32$ and $\alpha_s (\mu_H) \simeq 0.36$ of the QCD coupling constant.  Notice that the Wilson coefficient $C_{3 \tilde G}$ of the dimension-six  Weinberg operator gets strongly suppressed by one-loop~RG running in QCD.\footnote{The two-loop and three-loop  $N_F$-independent contributions to the  AD of the dimension-six Weinberg operator have been calculated very recently~\cite{deVries:2019nsu}.  Due to cancellations between the next-to-leading-logarithmic  and the next-to-next-to-leading-logarithmic QCD corrections, the total three-loop result is  numerically close to the one-loop result for $C_{3 \tilde G} (\mu_H)$ reported in~(\ref{eq:3GRG}). In view of this and given the sizeable uncertainties of the hadronic  matrix element of the dimension-six Weinberg operator $\big($cf.~(\ref{eq:3Gestimate})$\big)$ using only the LL RG evolution is in our opinion fully justified. In the same spirit, the two-loop and three-loop mixing of the quark EDMs and CEDMs~\cite{Misiak:1994zw,Gracey:2000am,Gorbahn:2005sa,Degrassi:2005zd} is also neglected in~(\ref{eq:3GRG}).} 

The hadronic  matrix elements of the dimension-six operators corresponding to the Wilson coefficients $d_q$, $\tilde d_q$ and $C_{3 \tilde G}$ in~(\ref{eq:3GRG})  are known with varying levels of theoretical uncertainties.  The EDM contributions from down and up quarks have been calculated  with an accuracy of $5\%$ using lattice QCD~(LQCD)~\cite{Bhattacharya:2015esa,Bhattacharya:2015wna,Gupta:2018lv}, while QCD sum-rule calculations~\cite{Pospelov:2000bw,Lebedev:2004va,Pospelov:2005pr,Hisano:2012sc} allow to determine the contributions from the down-quark and up-quark CEDMs with  uncertainties of $50\%$.  At present only estimates of the hadronic  matrix element of the dimension-six Weinberg operator exist  that rely on either QCD sum rules~\cite{Demir:2002gg,Haisch:2019bml}, the vacuum insertion approximation~\cite{Bigi:1990kz} or  naive dimensional analysis~\cite{Weinberg:1989dx}.  Since only the QCD sum-rules calculations allow for a systematic analysis of theoretical uncertainties, we will in the following rely on them. Adopting our very recent QCD sum-rule estimate~\cite{Haisch:2019bml}, which  is plagued by an uncertainty of~$50\%$, and  employing  $m_d (\mu_H) = 5.4 \cdot 10^{-3} \, {\rm GeV}$ and  $m_u (\mu_H) = 2.5 \cdot 10^{-3} \, {\rm GeV}$~\cite{Tanabashi:2018oca}, we obtain 
\beq \label{eq:3Gestimate} 
\frac{\big ( d_n \big )_{3 \tilde G} }{e} = \Big  [   \hspace{0.5mm} 1.0 \, \big ( 1 \pm 0.05 \big ) + 8.8 \, \big ( 1 \pm 0.5 \big ) - 66.6 \, \big ( 1 \pm 0.5 \big )  \hspace{0.5mm}  \Big ] \;     C_{3 \tilde G} (m_h)  \cdot 10^{-4} \, {\rm GeV}  \,,
\eeq
where the  first, second and third term corresponds to the $d_q$, $\tilde d_q$ and $C_{3 \tilde G}$ contribution in~(\ref{eq:3GRG}), respectively. We emphasise that in the case that   new physics  enters only through   the matching correction $C_{3 \tilde G} (m_h)$  the relative signs in~(\ref{eq:3Gestimate}) are all fixed, meaning that the  contribution from the dimension-six Weinberg operator necessarily interferes destructively with both the EDM and CEDM contribution. 

\subsection{Dimension-eight contribution}
\label{sec:dim8}

At LO the matching correction to  the Wilson coefficient $C_{4 \tilde G,1}$ proportional to $C_{\phi \tilde G}$ arise from one-loop graphs of the type  shown on the right in Figure~\ref{fig:diagrams2}. A straightforward calculation gives 
\beq \label{eq:4Gmatching} 
C_{4 \tilde G, 1} (m_h) =  \frac{\alpha_s (m_h)}{\pi} \frac{1}{m_h^2} \left  [ 
1 
+ \frac{7}{120 x} 
+  \frac{1}{168 x^2} 
\right ] C_{\phi \tilde G} (m_h) \,, 
\eeq
at the matching scale $\mu = m_h$.  Higher-order terms in the $1/x$ expansion  change the matching correction  $C_{4 \tilde G,1} (m_h)$  by less than a permille, and we  have therefore not included them in~(\ref{eq:4Gmatching}).\footnote{The  results for the ${\cal O} (1/x^3)$ and ${\cal O} (1/x^4)$ terms of~(\ref{eq:4Gmatching}) can be found in the {\LaTeX} source code of this paper.}  

At the dimension-eight level there are three independent CP-violating operators that can be built from QCD field strength tensors  (see~\cite{Morozov:1985ef,Chang:1991ry,Booth:1992iz} for details). While only the operator appearing  in~(\ref{eq:LCPVQCD}) receives a one-loop matching correction proportional to $C_{\phi \tilde G}$ all three dimension-eight operators mix under QCD. We write the two additional CP-violating four-gluon operators as 
\beq \label{eq:LW2}
 {\cal L}_{W2} =   - \frac{g_s^2}{12} \,  \tilde G_{\mu \nu}^a G^{b \, \mu \nu}  G_{\rho \lambda}^a G^{b \, \rho \lambda}  \,  C_{4 \tilde G, 2}  - \frac{g_s^2}{12} \,  d_{abe} \hspace{0.25mm} d_{cde}  \,   \tilde G_{\mu \nu}^a G^{b \, \mu \nu}  G_{\rho \lambda}^c G^{d \, \rho \lambda}  \,  C_{4 \tilde G, 3} \,,
\eeq
where $d_{abc}$ are totally symmetric structure constants of QCD. In the basis $\vec C_8 = \big (C_{4 \tilde G, 1} , C_{4 \tilde G,2} ,   C_{4 \tilde G, 3} \big)^T$  the one-loop AD matrix then reads~\cite{Morozov:1985ef,Chang:1991ry,Booth:1992iz}
\beq \label{eq:ADM4G} 
\hat \gamma_8 = \begin{pmatrix} \; \displaystyle -56 + \frac{8}{3} \hspace{0.25mm} N_F + 2 \hspace{0.25mm} \beta_0  \; &  \; 24  \; &   \; -36  \; \\[3mm]  \; -38  \; &  \;  \displaystyle 56 + \frac{8}{3} \hspace{0.25mm} N_F + 2 \hspace{0.25mm} \beta_0   \; &   \; -42  \; \\[3mm] \; -14  \; &  \; 12  \; &  \;  \displaystyle -14  + \frac{8}{3} \hspace{0.25mm} N_F + 2 \hspace{0.25mm} \beta_0  \; \end{pmatrix} \,.
\eeq
Working in the five, four and three flavour theory and using the values of the QCD coupling constant given earlier leads to the LL approximations
\beq \label{eq:RGdim8}
\begin{split}
C_{4 \tilde G, 1} (\mu_H) & \simeq 6.6 \hspace{0.5mm} C_{4 \tilde G, 1} (m_h) \,, \\[2mm]
C_{4 \tilde G, 2} (\mu_H) & \simeq -1.9 \hspace{0.5mm} C_{4 \tilde G, 1} (m_h) \,, \\[2mm]
C_{4 \tilde G, 3} (\mu_H) & \simeq 3.9 \hspace{0.5mm} C_{4 \tilde G, 1} (m_h) \,.
\end{split}
\eeq
Notice that in contrast to~(\ref{eq:3GRG}) the Wilson coefficient $C_{4 \tilde G, 1}$ of the dimension-eight Weinberg operator that is generated at the matching scale gets enhanced by RG running. 

Estimates of the  hadronic  matrix elements of the dimension-eight  CP-violating four-gluon operators  have very recently been obtained in the context of QCD sum rules~\cite{Haisch:2019bml}. Employing these results we find  
\beq \label{eq:4Gestimate} 
\frac{\big ( d_n \big )_{4 \tilde G} }{e} =  -6.1 \,  \big ( 1 \pm 0.8 \big ) \, C_{4 \tilde G, 1} (m_h) \cdot 10^{-1} \,  {\rm GeV}^3  \,,
\eeq
which has a theoretical uncertainty of 80\%. We add that the sign in $(d_n)_{4 \tilde G}/e$ is predicted in the QCD sum-rule approach and that neglecting all contributions from the two additional dimension-eight operators~(\ref{eq:LW2}) would lead to a numerical result that deviates from~(\ref{eq:4Gestimate}) by less than~5\%. 

\section{Discussion}
\label{sec:discussion}

To discuss the constraints that nEDM measurements can set on the CP-violating dimension-six Higgs-gluon interactions appearing in~(\ref{eq:LCPVQCD}),  we introduce the dimensionless Wilson coefficient 
\beq \label{eq:dimlessWC}
 \bar C_{\phi \tilde G} (m_h)  = v^2 \, C_{\phi \tilde G} (m_h) \,.
\eeq
Combining (\ref{eq:3Gestimate}) and (\ref{eq:4Gestimate}) we then find  in terms of the bared  Wilson coefficient the following expression
\beq \label{eq:dnestimateall} 
\left | \frac{d_n   }{e}  \right  |   =  6.2 \, \big  | \left (1 \pm 0.05 \right) + 8.6 \left (1 \pm 0.5 \right) - 65.1 \left (1 \pm 0.5 \right)  -   7.5  \left (1 \pm 0.8 \right)  \big  |  \, \left | \bar C_{\phi \tilde G} (m_h) \right | \cdot 10^{-26} \, {\rm cm} \,, 
\eeq
where the contributions associated to the terms $d_q$, $\tilde d_q$, $C_{3 \tilde G}$ and the dimension-eight Weinberg operators have been  kept distinct. 

In view of the sizeable hadronic uncertainties of the  matrix elements of the operators of CEDM and Weinberg type and the relative overall sign of the Weinberg-type contributions, we combine the errors  in~(\ref{eq:dnestimateall}) in such a way that our prediction
\beq \label{eq:dnestimate} 
\left | \frac{d_n}{e}  \right  | = 1.3 \, \left | \bar C_{\phi \tilde G} (m_h) \right | \cdot 10^{-24} \, {\rm cm} 
\eeq
provides a lower absolute limit on the actual size of the $\bar  C_{\phi \tilde G}$ corrections to $d_n$. Because our error treatment assumes a cancellation between the numerically  dominant contributions associated to $C_{3 \tilde G}$, $\tilde d_q$ and the dimension-eight terms, it is conservative.  Notice that if such a cancellation is not at work in practice, predictions for $\big | \hspace{0.125mm} d_n/e \hspace{0.125mm}  \big  |$  can be obtained that are larger by a factor of about 5 than the upper limit~(\ref{eq:dnestimate}) but still consistent within the individual uncertainties quoted  in~(\ref{eq:dnestimateall}). 

The current experimental nEDM results~\cite{Baker:2006ts,Afach:2015sja} impose the following 95\%~CL bound 
\beq \label{eq:presentCL95}
\left | \frac{d_n   }{e}  \right  | <  3.6 \cdot 10^{-26} \, {\rm cm} \,, 
\eeq
which, using~the lower limit of the only quark EDMs case in~(\ref{eq:dnestimate}) translates into 
\beq \label{eq:CphiGcurrentnEDM}
\left | \bar C_{\phi \tilde G} (m_h) \right | < 2.9 \cdot 10^{-2} \,.
\eeq
 As shown in Appendix~\ref{sec:SMyukawas}, this bound is weaker by a factor of almost 30 than the 95\%~CL exclusion obtained in universal theories that assume that the Yukawa couplings of the light quarks are SM-like. The constraint~(\ref{eq:CphiGcurrentnEDM}) can also be compared to the 95\%~CL limit
\beq \label{eq:CphiGcurrentLHC}
\bar C_{\phi \tilde G} (m_h) \in \left [-0.13, 0.83 \right ]  \cdot 10^{-2} 
\eeq
on the dimensionless Wilson coefficient~(\ref{eq:dimlessWC}) that has been obtained in~\cite{Bernlochner:2018opw} from an analysis of the azimuthal angle difference $\Delta \phi_{jj}$  between the two jets in $h + 2 j$ LHC events. 

In order to obtain an idea of the prospects of the low-energy constraints, we assume that a lower bound of 
\beq \label{eq:futureCL95}
\left | \frac{d_n   }{e}  \right  | <  1.0 \cdot 10^{-27} \, {\rm cm} 
\eeq
can be set at  the proposed PSI and LANL nEDM experiments~\cite{Schmidt-Wellenburg:2016nfv,Ito:2017ywc}. In such a case, one arrives at the  limit 
\beq \label{eq:CphiGfuturenEDM}
\left | \bar C_{\phi \tilde G} (m_h) \right | < 8.0 \cdot 10^{-4} \,,
\eeq
if one assumes  that~(\ref{eq:dnestimate}) provides a lower absolute limit on the  $\bar  C_{\phi \tilde G}$ corrections to $d_n$. The sensitivity study~\cite{Bernlochner:2018opw}  of the $pp \to h + 2 j$ process finds on the other hand that the high-luminosity LHC~(HL-LHC)  should be able to set a bound of 
\beq \label{eq:CphiGfutureLHC}
\left | \bar C_{\phi \tilde G} (m_h) \right | <  9.2 \cdot 10^{-4} \,.  
\eeq

A comparison of the limits~(\ref{eq:CphiGcurrentnEDM}) and (\ref{eq:CphiGcurrentLHC})  on the CP-violating interactions involving the Higgs boson and gluons shows that at present the sensitivity of nEDM searches is by roughly an order of magnitude weaker than the constraining power of the LHC. This is mainly a result of the conservative treatment  of  the hadronic uncertainties in~(\ref{eq:dnestimateall}) that led to~(\ref{eq:dnestimate}). From (\ref{eq:CphiGfuturenEDM}) and (\ref{eq:CphiGfutureLHC}) it is however also evident that in the future the sensitivity of nEDM searches can reach the LHC level  even if the accuracy  of the hadronic matrix elements are not improved.  First-principle calculations of the matrix elements of the CEDMs and the dimension-six Weinberg operator  are possible using existing  LQCD methodology, and considering the efforts by several LQCD groups~\cite{Bhattacharya:2015rsa,Bhattacharya:2016rrc,Abramczyk:2017oxr,Dragos:2017wms,Rizik:2018lrz,Kim:2018rce,Bhattacharya:2018qat,Syritsyn:2019vvt},  it seems possible that estimates with uncertainties similar to  the current ones can be obtained within the next five years~\cite{Gupta:2019fex,Cirigliano:2019jig}. It remains to be seen which accuracy such computations can achieve in the next 20 years of LHC running, but we believe that it is very likely that the bound~(\ref{eq:CphiGfuturenEDM}) can  be improved by the end of the HL-LHC run. 

From the above discussion one can conclude that future nEDM searches and LHC measurements are complementary to each other even in the specific class of new-physics models where the Yukawa couplings of light quarks are zero. Since such new-physics realisations represent in some sense the worst-case scenario for the low-energy constraints considered here, we  believe this to be an interesting finding. Our results show furthermore that in a global SMEFT analysis, EDM constraints can have additional flat or weakly bound directions that do not appear in the case of the class of universal theories considered in~\cite{Panico:2018hal,Cirigliano:2019vfc}.  As already emphasised in~\cite{Cirigliano:2019vfc}, to resolve unbounded directions in the multi-dimensional space of Wilson coefficients, high-$p_T$ and low-energy constraints on CP-violating couplings between the Higgs and gauge bosons should be combined into global fits. The~nEDM results presented in this article  can be readily used for such a purpose. 
 
\acknowledgments The Feynman diagrams shown in this article and the  analytic expression for the relevant one-loop and two-loop scattering amplitudes have been obtained using {\tt FeynArts}~\cite{Hahn:2000kx}.  The  Feynman~rules for the higher-dimensional operators~(\ref{eq:LCPVexample}) and (\ref{eq:LCPVQCD}) have been derived with the help of the~{\tt FeynRules} package~\cite{Alloul:2013bka}. 

\appendix

\section{Case of universal theories}
\label{sec:SMyukawas}

In this appendix we extend the formalism presented in Section~\ref{sec:dim6} to the case of universal theories. Under the assumption that the  light-quark Yukawa couplings take the values predicted in the SM (i.e.~$\kappa_q = 1$) the one-loop diagram on the  right-hand side in Figure~\ref{fig:diagrams1} induces a CEDM for the down and up quark. In agreement with~\cite{Cirigliano:2019vfc}, we find  the following expression for the one-loop correction to  the CEDMs
\beq \label{eq:CEDMmatching}
\tilde d_q (m_h) = \frac{3  \alpha_s (m_h)}{2 \hspace{0.125mm} \pi} \, C_{\phi \tilde G} (m_h) \,,
\eeq
at the matching scale~$\mu = m_h$. Including the contribution~(\ref{eq:CEDMmatching}) in the evaluation of the nEDM, the formula~(\ref{eq:3Gestimate}) turns into 
\bea \label{eq:3GestimateSMlike} 
\begin{split}
\frac{\big ( d_n \big )_{\tilde d_q, 3 \tilde G} }{e} = \Big  [   \hspace{0.5mm} & \left ( 6.9  \hspace{0.5mm} \tilde d_q (m_h)  + 1.0  \hspace{0.5mm}  C_{3 \tilde G} (m_h)  \right ) \, \big ( 1 \pm 0.05 \big )  +\left ( 36.7  \hspace{0.5mm} \tilde d_q (m_h) + 8.8 \hspace{0.5mm}   C_{3 \tilde G} (m_h)  \right ) \, \big ( 1 \pm 0.5 \big ) \hspace{8mm} \\[2mm] & - 66.6 \hspace{0.5mm}  C_{3 \tilde G} (m_h)  \, \big ( 1 \pm 0.5 \big )    \hspace{0.5mm}  \Big ] \;   \cdot 10^{-4} \, {\rm GeV}  \,.
\end{split}
\eea

Adding the new dimension-six contribution~(\ref{eq:3GestimateSMlike}) to the dimension-eight piece~(\ref{eq:4Gestimate}) we then obtain the  formula 
\beq \label{eq:dnestimateallSMlike} 
\begin{split}
\left | \frac{d_n   }{e}  \right  |   =  6.2 \, \big  |  \hspace{0.25mm} &190.6 \left (1 \pm 0.05 \right) + 1016.6 \left (1 \pm 0.5 \right) \\[2mm] 
& - 65.1 \left (1 \pm 0.5 \right)  -   7.5  \left (1 \pm 0.8 \right)  \big  |  \, \left | \bar C_{\phi \tilde G} (m_h) \right | \cdot 10^{-26} \, {\rm cm} \,, 
\end{split}
\eeq
where $\bar C_{\phi \tilde G}$ denotes the dimensionless Wilson coefficient introduced in~(\ref{eq:dimlessWC}). In view of the sizeable hadronic uncertainties of the  matrix elements in~(\ref{eq:dnestimateallSMlike}) and the relative overall sign of the Weinberg-type contributions, we again combine the errors  in $\big | \hspace{0.125mm} d_n/e \hspace{0.125mm}  \big  |$ in such a way that our final prediction provides a lower absolute limit on the actual size of the $\bar  C_{\phi \tilde G}$ corrections to the nEDM. We find  
\beq \label{eq:dnestimateSMlike} 
\left | \frac{d_n}{e}  \right  | = 3.6 \, \left | \bar C_{\phi \tilde G} (m_h) \right | \cdot 10^{-23} \, {\rm cm} \,.
\eeq

Numerically, the result~(\ref{eq:dnestimateSMlike}) implies that 
\beq \label{eq:CphiGnEDMSMlike}
\begin{split}
& \left | \frac{d_n   }{e}  \right  | <  3.6 \cdot 10^{-26} \, {\rm cm}  \; \; \; \Rightarrow \; \; \; \left | \bar C_{\phi \tilde G} (m_h) \right | < 1.0 \cdot 10^{-3} \,, \\[2mm]
& \left | \frac{d_n   }{e}  \right  | <  1.0 \cdot 10^{-27} \, {\rm cm}  \; \; \; \Rightarrow \; \; \; \left | \bar C_{\phi \tilde G} (m_h) \right | < 2.8 \cdot 10^{-5} \,.
\end{split}
\eeq
Notice that the 95\%~CL limits on $ \left | \bar C_{\phi \tilde G} (m_h) \right |$ as given in~(\ref{eq:CphiGnEDMSMlike}) are comparable to the bounds that have been derived in~\cite{Cirigliano:2019vfc} by using the so-called Rfit strategy, in which all hadronic matrix elements entering the prediction for the nEDM are varied within their allowed ranges. 


%

\end{document}